\documentclass{PoS}
\usepackage{url}

\title{Status of the MUonE experimental proposal}

\ShortTitle{Status of the MUonE experimental proposal}

\author{\speaker{G. Venanzoni}\thanks{For the MUonE Collaboration}\\
       INFN Sezione di Pisa, Pisa, Italy\\
        E-mail: \email{graziano.venanzoni@pi.infn.it}}

\abstract{We present the status of the MUonE experimental proposal which aims
  at determining the leading order hadronic contribution to the muon g-2 by measuring the hadronic part of the photon vacuum polarization in the space-like region.}
 
\FullConference{
European Physical Society Conference on High Energy Physics - EPS-HEP2019 -\\
			10-17 July, 2019\\
			Ghent, Belgium}

\begin{document}

\section{Introduction}
There is a tantalizing discrepancy of $\sim 3.5$ standard deviations between the measurement of the muon anomaly $a_{\mu}=(g-2)/2$ performed by the E821 experiment at BNL and the Standard Model prediction~\cite{Blum:2013xva}. Whether this discrepancy is real or not, it certainly calls for a more precise determination of $a_{\mu}$.
New experiments at Fermilab (E989, an evolution of E821) and at J-PARC (E34, with a completely different technique) aim to measure $a_{\mu}$ to 0.14 ppm.
%Luckily enough a new muon g-2 experiment is undergoing at Fermilab (E989, an ev olution of E821), and another one (E34, with a completely different technique) is under development at J-PARC. They both plan to improve the experimental uncertainty by a factor of 4.
With the planned improvement of the measurement, it is important that the theoretical prediction improves as well.
The leading-order hadronic vacuum polarization contribution, $a_{\mu}^{\rm{HLO}}$,
currently represents the main limitation for the theory due to the non-perturbative QCD behavior at low energy. An intense research program is underway with both time-like data and lattice calculations~\cite{TH}.
%%%%%%%%%%%%%%%%%%%%%%%%%%%%%%%%%%%%%%%%%%%%%%%%%%%%
\section{The MUonE Project}
%%%%%%%%%%%%%%%%%%%%%%%%%%%%%%%%%%%%%%%%%%%%%%%%%%%%

A novel approach has been proposed recently to determine the leading hadronic contribution to the muon $g$-2 ($a_\mu^{\rm HLO}$) measuring the effective electromagnetic coupling in the space-like region via scattering data~\cite{Calame:2015fva}. The elastic scattering of high-energy muons on atomic electrons of a low-$Z$ target has been identified as an ideal process for this measurement, and a new experiment, MUonE, has been proposed at CERN to measure the shape of the differential cross section of $\mu e$ elastic scattering as a function of the space-like squared momentum transfer~\cite{Abbiendi:2016xup}. 
Assuming a 150~GeV muon beam with an average intensity of $\sim 1.3 \times 10^7$ muons/s, presently available at the CERN muon M2 beamline, incident on a target consisting of 40 beryllium layers, each 1.5~cm thick, and three years of data taking, one can reach an integrated luminosity of about $1.5 \times 10^7~{\rm nb}^{-1}$, which would correspond to a statistical error of $0.3\%$ on the value of $a_\mu^{\rm HLO}$. The direct measurement of the effective electromagnetic coupling via $\mu e$ scattering would therefore provide an independent and competitive determination of $a_\mu^{ \rm HLO}$. It would consolidate the muon $g$-2 prediction and allow a firmer interpretation of the upcoming measurements at Fermilab and J-PARC.

%%%%%%%%%%%%%%%%%%%%%%%%%%%%%%%%%%%%%%%%%%%%%%%%%%%%
\subsection{The experiment}
%%%%%%%%%%%%%%%%%%%%%%%%%%%%%%%%%%%%%%%%%%%%%%%%%%%%

\begin{figure*}[htp]
\includegraphics[width=.5\textwidth]{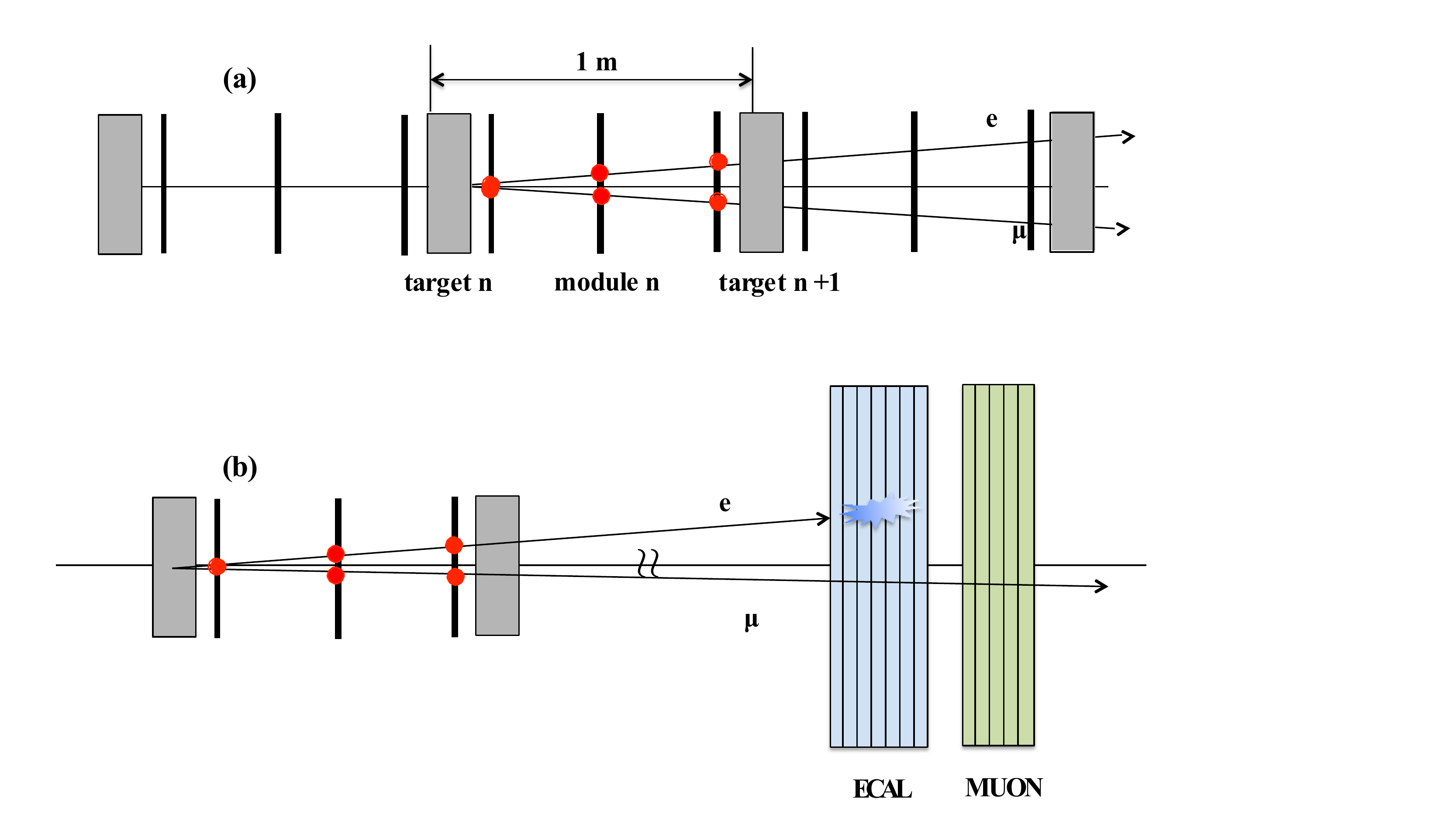}~\includegraphics[width=.5\textwidth]{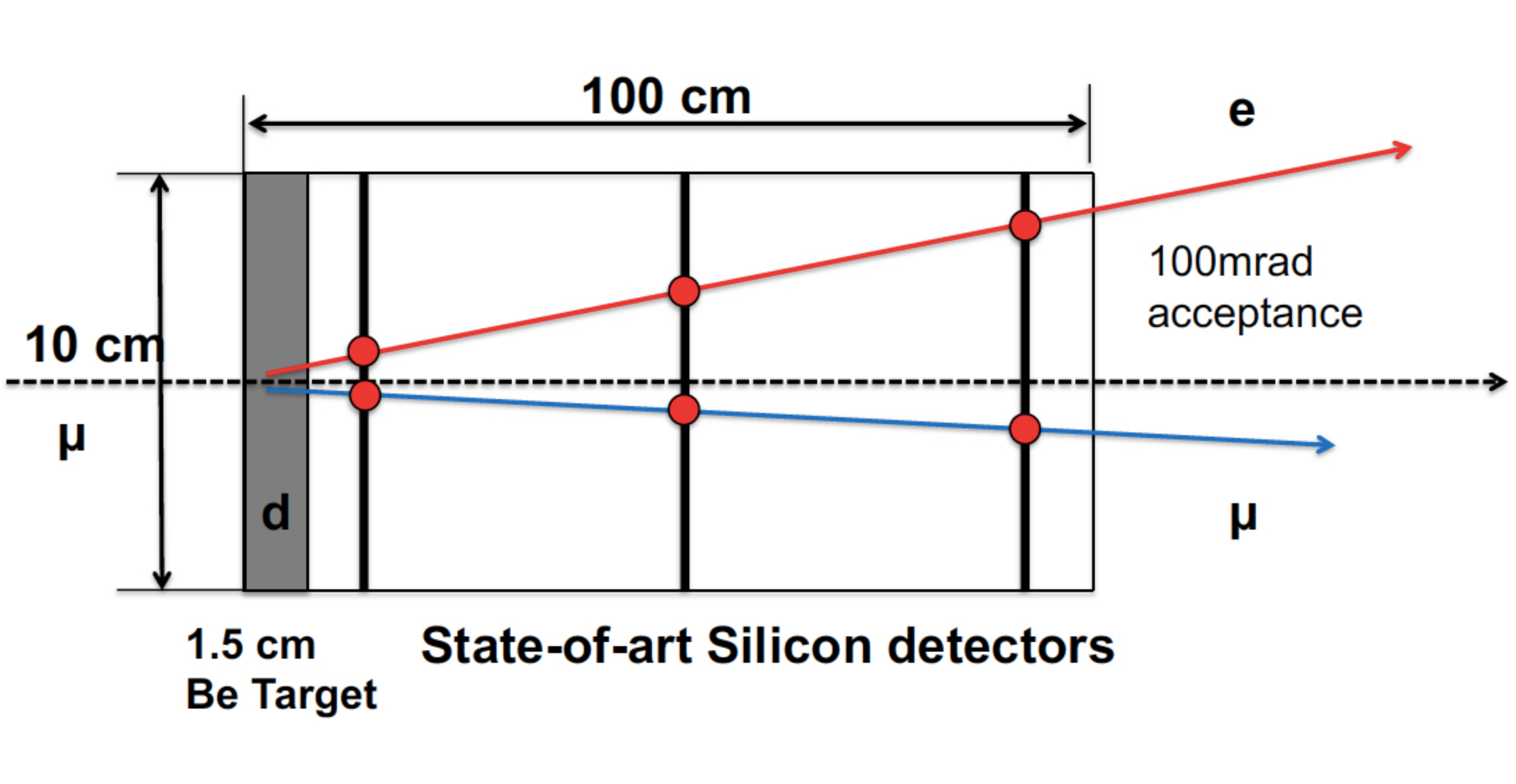}
\caption{Left: design of the baseline detector concept. Right: single unit.}
  \label{Be-detector}
\end{figure*}
Figure~\ref{Be-detector} shows the baseline detector design. The detector is a repetition of 40 identical modules (called stations), each consisting of a 1.5~cm thick layer of Be coupled to 3 silicon tracking layers within a distance of $\sim1$~m (to be optimized) from each other with intermediate air gaps.  Thin targets are required to minimize the impact of multiple scattering and the background on the measurement. Several targets allow to obtain the necessary statistics. The Si detectors provide the necessary resolution ($\sim$20 $\mu$m) with a limited material budget ($< 0.06 \, X_0$ per unit). This arrangement provides both a distributed target with low-{\it Z} and the tracking system. Downstream of the apparatus a calorimeter and a muon system (a filter plus active planes) will be used for e/$\mu$ particle identification. Significant contributions of the hadronic vacuum polarization to the $\mu e \to \mu e$ differential cross section are essentially restricted to electron scattering angles below 10~mrad, corresponding to electron energies above 10~GeV. The net effect of these contributions is to increase the cross section by a few per mille: a precise determination of $a_{\mu}^{\rm{HLO}}$ requires not only high statistics, but also a high systematic accuracy, as the final goal of the experiment is equivalent to a determination of the signal to normalization ratio with a $O(10~\rm ppm)$ systematic uncertainty at the peak of the integrand function. Although this does not require knowledge of the absolute cross section (signal and normalization regions will be obtained by $\mu e$ data) it poses severe requirements on the knowledge of the following quantities:

\begin{itemize}
\item {Multiple scattering:} preliminary studies indicate that an accuracy of the order of $\sim$ 1\% is required on the knowledge of the multiple scattering effects in the core region. Results from a Test Beam at CERN with electrons of 12 and 20~GeV on 8-20~mm C target show good agreement between data and GEANT4 simulations, see Fig.~\ref{ms}~\cite{Abbiendi:2019qtw}.

\begin{figure*}[htp]
\includegraphics[width=.5\textwidth]{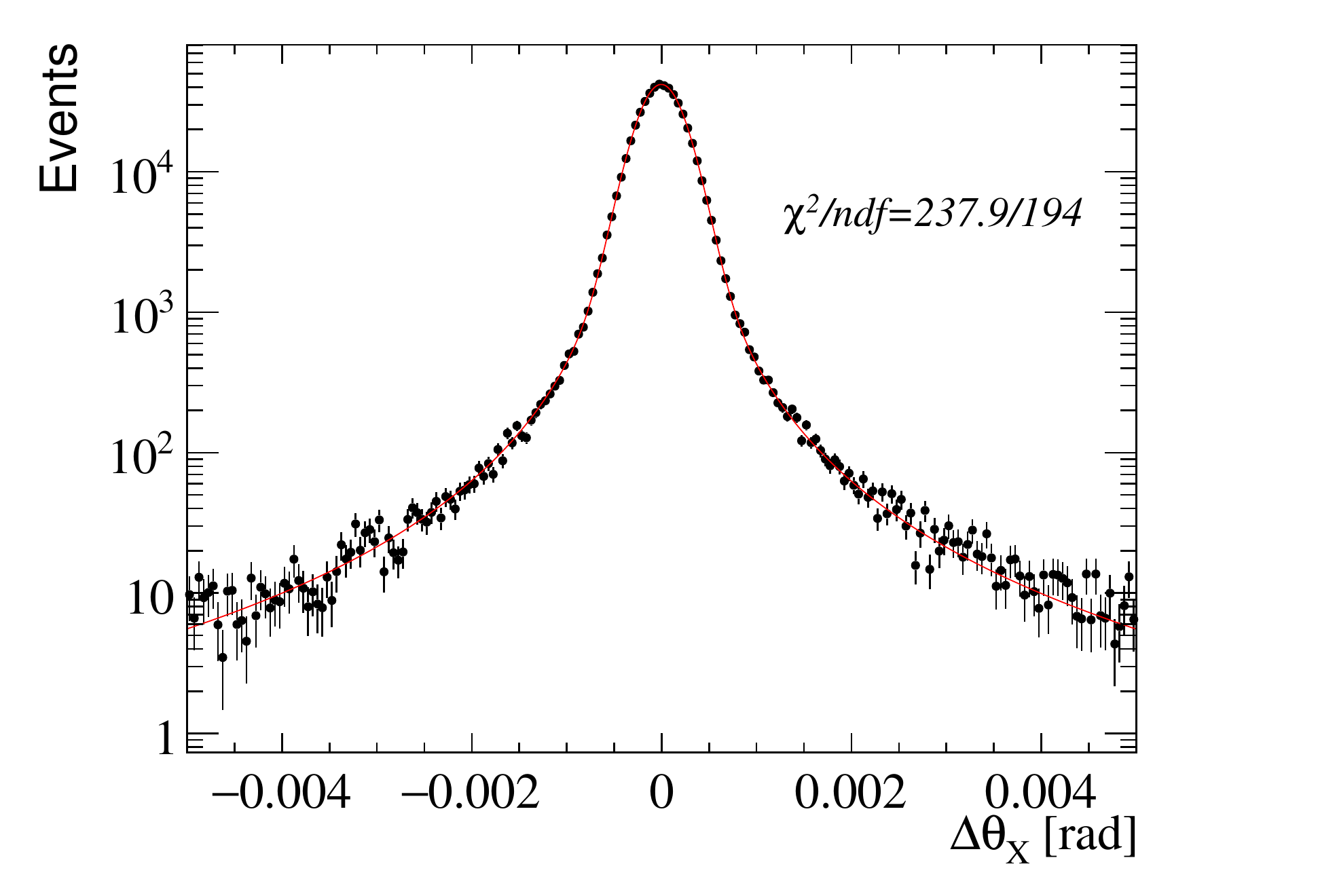}~\includegraphics[width=.5\textwidth]{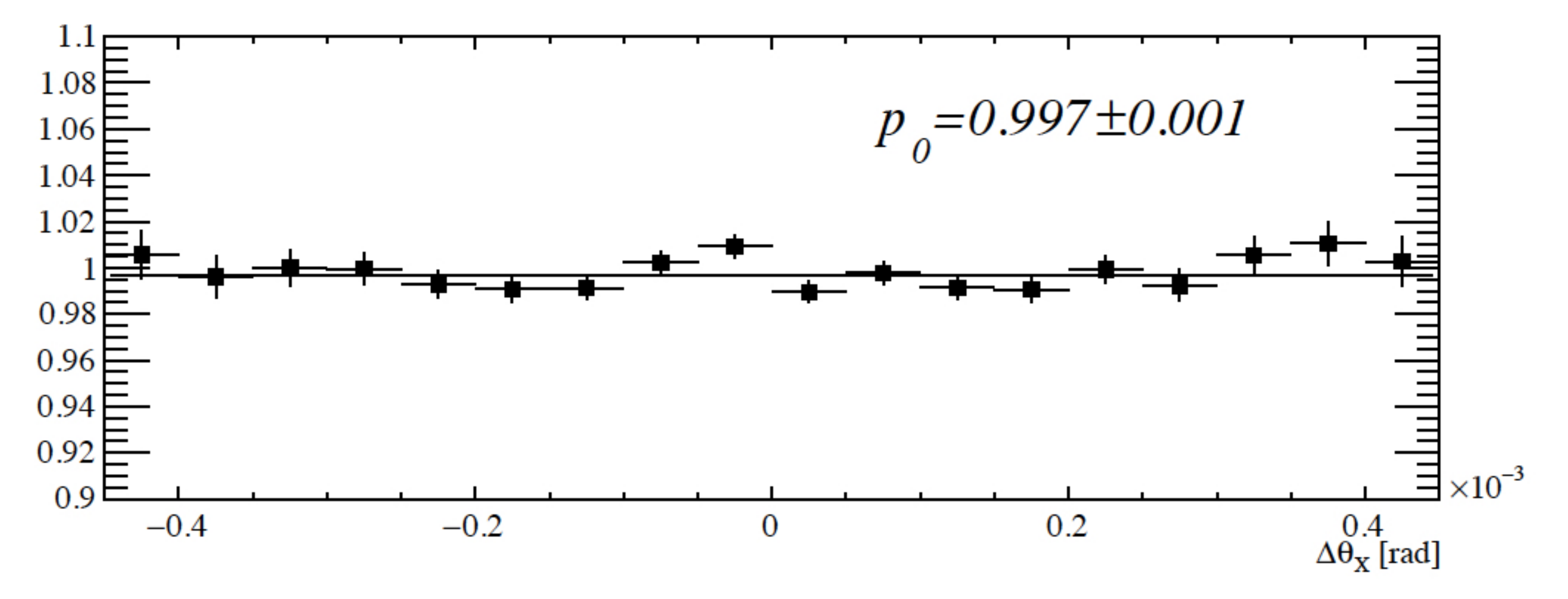}
\caption{Left: x-projection of the scattering angle from 12 GeV $e^-$ with 8 mm target compared with the results of the fit based on the sum of a
Gaussian and a Student's $t$ distribution. Right: Data/Monte Carlo ratio.}
  \label{ms}
\end{figure*}

\item{Tracking uniformity, alignment and reconstruction of angles:} it is important to keep the systematic error arising from the non-uniformity of the tracking efficiency and angle reconstruction at the 10$^{-5}$ level. The use of state-of-the-art silicon detectors should ensure the required uniformity. 
  Among the considered alternatives, the silicon strip sensors being developed for the CMS Tracker upgrade represent a good solution. In particular, the silicon sensors which are foreseen for the CMS HL-LHC Outer Tracker (OT) in the so called 2S configuration have been chosen \cite{cmsu}. They are 320 $\mu$m thick sensors with n-in-p polarity produced by Hamamatsu Photonics. They have an area of 10 cm$\times$10 cm (sufficient to cover the MUonE acceptance) and a pitch p = 90 $\mu$m, which means having a single hit precision $\sim p/
  \sqrt{12} \sim 26 \mu$m. The strips are capacitively-coupled, and are segmented in two approximately 5 cm long strips. In the 2S configuration two closely-spaced silicon sensors reading the same coordinate are mounted together and read out by common front-end ASIC.  With their accompanying
front end electronics they can sustain high readout rate (40 MHz) and are well-suited for track triggering.

%  Each sensor features a large active area (sufficient to cover the MUonE acceptance) and an adequate spatial resolution. It can sustain the high readout rate required for MUonE (40 MHz) with their accompanying
%front-end electronics. In the so-called 2S configuration they are well-suited for track triggering. They
%also represent a good compromise for detector thickness. Their production schedule is also in line with
%the timescale of MUonE, as described in Section 14, and they have been chosen for the experiment.
 % Their production is also in line with the MUonE timescale. In particular, the silicon sensors which are foreseen for the CMS HL-LHC Outer Tracker (OT) in the so called 2S configuration have been chosen [48]. They are 320 $\mu$m thick sensors with n-in-p polarity produced by Hamamatsu Photonics. They have a squared area of 10 cm$\times$10 cm and a pitch p = 90 $\mu$m, which means having a single hit precision $\sim p/
 % sqrt(12) \sim 26 \mu$m. The strips are capacitively-coupled, and are segmented in two approximately 5 cm long strips. 

  \begin{figure*}[htp]
    \begin{center}
\includegraphics[width=1.0\textwidth]{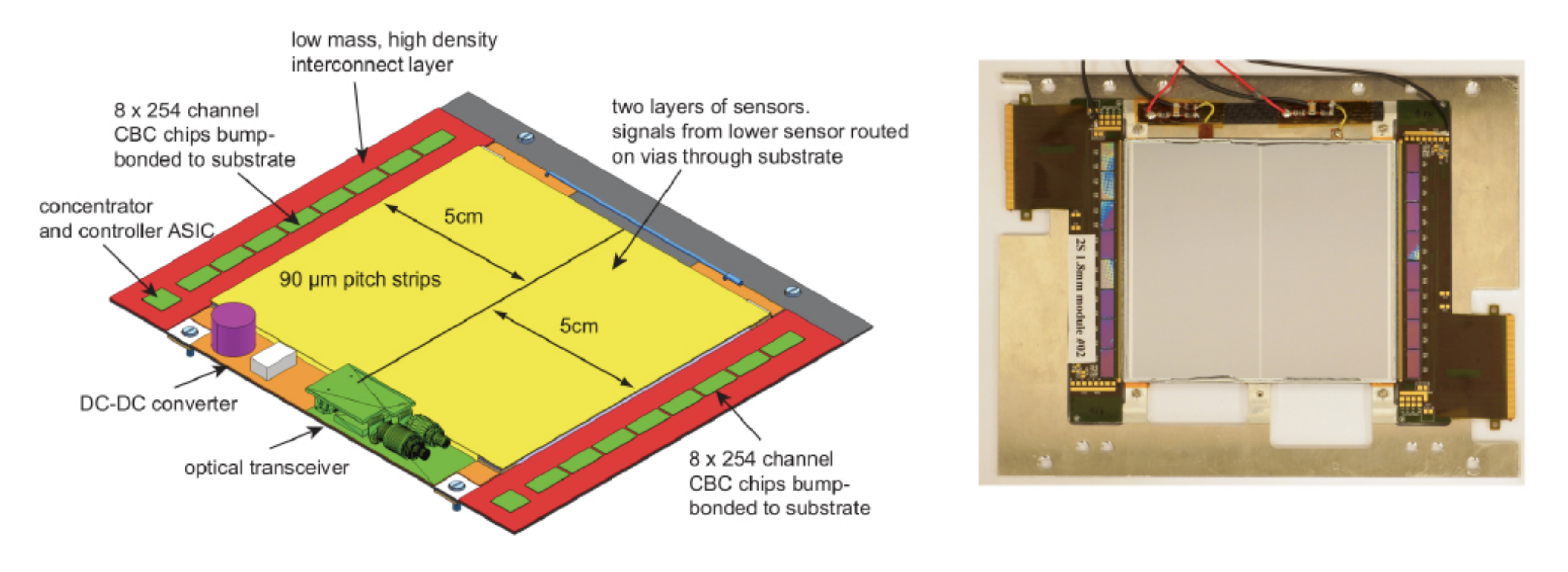}
\caption{Left: A schematic view of the CMS 2S module.  Right: A picture of the CMS 2S module.}
\label{cms}
\end{center}
\end{figure*}

  The relative alignment of the silicon detectors will be monitored with the high statistics provided by the muon beam. The longitudinal position of the silicon detector must be controlled at the level of 10 $\mu$m.
  The support structure able to meet this requirement is shown in Fig.~\ref{support}.
  The base is a U shaped carbon fiber
  structure; the vertical dimension - 20 mm in figure - guarantees stiffness to vertical sag.  Carbon fiber has a negligible coefficient of thermal expansion
  (CTE)
in the longitudinal direction, which can be theoretically reduced to zero in the region of interest (T$\simeq$ 300 K). Moreover its density d=1.42 g/cm$^3$ is very low, thus limiting the backscattering effect.
The relative distance between the Si tracker elements will be monitored by a laser-interferometry system. 

\begin{figure*}[htp]
\begin{center}
  \includegraphics[width=.5\textwidth]{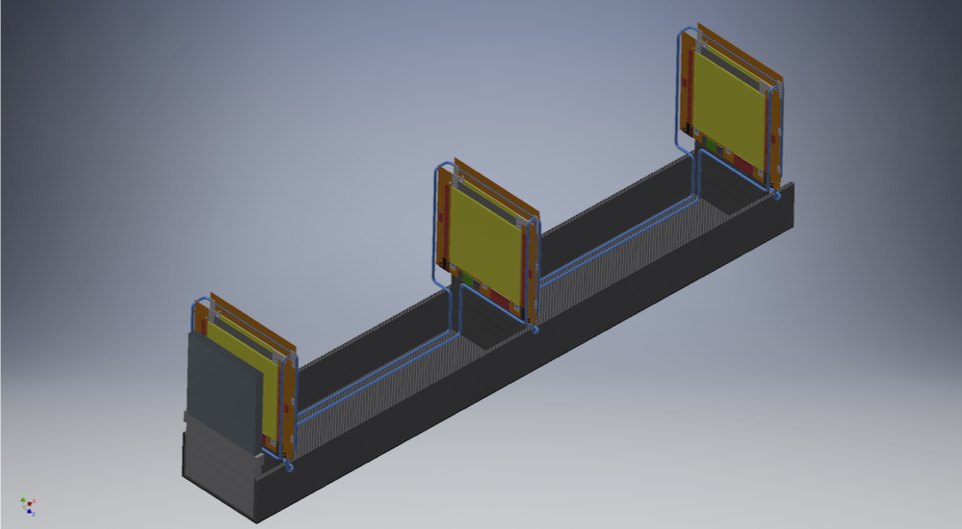}
  \caption{Support structure in carbon fiber.}
  \label{support}
  \end{center}
\end{figure*}

\item{Knowledge of the beam:} a 0.8\% accuracy on the knowledge of the beam momentum, as obtained by the BMS spectrometer used by COMPASS, is sufficient to control the systematic effects arising from beam spread. The beam scale must be known at $\sim 5$~MeV level. This can be obtained by $\mu e$ elastic scattering events exploring the $\mu e$ kinematics~\cite{loi}.

\item{Extraction of $\Delta\alpha_{had}(t)$ in presence of NLO effects:}
  The signal extraction is carried out by a template fit method.
  $\Delta\alpha_{had}(t)$ is modeled by a two-parameter analytical function
 with logarithmic dependency at
large $|t|$ and linear behavior at small $|t|$, as expected from general principles~\cite{loi}.
  Template distributions for the scattering
angles $\theta_{\mu}$ and $\theta_e$, both 1D and 2D, have been calculated from NLO Monte Carlo events on a grid of
points in the parameter space sampling the region around the expected reference values.
The template fit is then carried out by a $\chi^2$ minimization, comparing the angular distribution of pseudodata
with the predictions obtained for the scanned grid points. Extraction of
$a_{\mu}^{HLO}$ is consistent with the expected value within half standard deviation.
\end{itemize}

%%%%%%%%%%%%%%%%%%%%%%%%%%%%%%%%%%%%%%%%%%%%%%%%%%%%
\subsection{Theory}
%%%%%%%%%%%%%%%%%%%%%%%%%%%%%%%%%%%%%%%%%%%%%%%%%%%%
%Next-to-Leading order (NLO) QED corrections to the differential cross section were computed long time ago by applying some approximations
%and revisited more recently in~\cite{Kaiser:2010zz}.
The complete calculation of the full set of NLO QED corrections and of NLO electroweak corrections with the development of a fully exclusive Monte Carlo event generator for MUonE was completed in~\cite{Alacevich:2018vez}. The generator is currently used for simulation of MUonE events in presence of QED radiation.
The QED corrections at next-to-next-to-leading order (NNLO), crucial to interpret MUonE high-precision data, are not yet known.
%although some of the two-loop corrections which were computed for Bhabha scattering in QED~\cite{Bern:2000ie,Bonciani:2003te,Bonciani:2003cj} and for $t{\bar t}$ production in QCD~\cite{Bonciani:2008az, Bonciani:2013ywa} can be applied to $\mu e$ scattering as well.
A first step towards the calculation of the full NNLO QED corrections to $\mu e$ scattering was taken in~\cite{Mastrolia:2017pfy,DiVita:2018nnh}, where the master integrals for the two-loop planar and non-planar four-point Feynman diagrams were computed.
%These integrals were calculated setting the electron mass to zero, while retaining full dependence on the muon one. The extraction of the leading electron mass effects from the massless $\mu e$ scattering amplitudes has been recently addressed in~\cite{Engel:2018fsb} (see also~\cite{Penin:2005kf,Mitov:2006xs,Becher:2007cu}). 
The NNLO hadronic corrections to $\mu e$ scattering have been computed very recently in~\cite{Fael:2018dmz,Fael:2019nsf}. A suitable subtraction scheme to deal with soft singularities at NNLO accuracy in QED, including finite fermion masses, has been presented in~\cite{Engel:2019nfw}.
%using the dispersive approach
%with hadronic $e^+e^-$ annihilation (timelike) data.
%This approach, originally based on~\cite{Cabibbo:1961sz}, has also been employed to
%calculate the hadronic corrections to muon decay~\cite{vanRitbergen:1998hn,Davydychev:2000ee} and Bhabha scattering~\cite{Actis:2007fs,Kuhn:2008zs,CarloniCalame:2011zq}. The results of~\cite{Fael:2019nsf} show that these corrections will play a crucial role in the analysis of MUonE data. Recently, taking advantage of the hyperspherical integration method, it was shown that these NNLO hadronic corrections can also be calculated employing the hadronic vacuum polarization in the spacelike region, without using timelike data~\cite{Fael:2018dmz}.
%
%To NNLO accuracy electron pair production effects should be also considered and taken into account. One and two loop diagrams with vacuum polarization insertions in the photon propagator were considered some time ago~\cite{Arbuzov:1995cn,Arbuzov:1995vi} for the case of the Bhabha scattering only in the massless limit up to the 0.1\% accuracy. Pair production shows in fact, potentially dangerous unrenormalizable singularities of the form $(\alpha/\pi)^2\log (-t/m_e^2 )^3$ which are only canceled by the real pair production diagrams at the same order. The resummation of lepton pair contributions has also been proven~\cite{Catani:1989et}.

The extreme accuracy of MUonE demands for the resummation of classes of radiative corrections which are potentially enhanced by large logarithms.
%RCs can be organized in a power series of $\alpha/(2\pi)$ times powers of $L\equiv\log\left(-t/m_e^2\right)$ (we refer here only to the case of radiation from the electron leg, which is numerically the most relevant one) and $\ell\equiv-2\log\left(2\Delta\omega/\sqrt{s}\right)$, where $L$ and $\ell$ are the so-called {\em collinear-log} and {\em IR-log} (or {\em soft-log}), respectively. In the definition of $\ell$, $\Delta\omega$ is related to the maximum energy allowed for the radiation, which is in general a function of the applied cuts and the observable under consideration. Thanks to factorization theorems of soft and collinear radiation, the resummation techniques exponentiate the leading-log corrections up to all orders in $\alpha$ (terms of the form $\alpha^n (L-1)^n l^n$).
A general framework for implementing numerically the leading logarithmic resummations is provided either by the parton shower (PS) approach or the YFS formalism. These methods can be improved to include consistently NLO corrections~\cite{Montagna:1996gw,Balossini:2006wc,Jadach:1995nk}. Going one step further, when the complete NNLO corrections will be available and a NNLO matched PS (or ${\cal O}(\alpha^2)$ YFS) will be implemented, we expect that the error due to missing corrections will be  ${\cal O}(10^{-6})$~\cite{loi}.
%will start at order $\alpha^3L^2$, not enhanced by any IR-log $\ell$; i.e., we expect that the theoretical error on any distribution will be of the order of
%$
%(\alpha/2\pi)^3(2L)^2\times k \simeq 10^{-6}\times k,
%$
%with $k$ of ${\cal O}(1)$.

%%%%%%%%%%%%%%%%%%%%%%%%%%%%%%%%%%%%%%%%%%%%%%%%%%%%
\subsection{Status and future plans}
%%%%%%%%%%%%%%%%%%%%%%%%%%%%%%%%%%%%%%%%%%%%%%%%%%%%
At present, the MUonE Collaboration consists of groups from CERN, China, Germany, Italy, Poland, Russia, Switzerland, UK, and USA. These groups have strong expertises in the field of precision physics. A Letter of Intent has been submitted in June 2019 to
CERNS SPSC~\cite{loi}. The year 2020 will be devoted to continuing detector optimization
studies, simulations, and theory improvement. The detector construction is expected during CERN LS2 and the plan is to have a first pilot run of a few weeks in 2021. A run at full statistics is envisaged in 2022--24. MUonE is part of the PBC Study Group at CERN~\cite{pbc}.

%%%%%%%%%%%%%%%%%%%%%%%%%%%%%%%%%%%%%%%%%%%%%%%%%%%%

\end{document}